\documentclass[11pt,twoside]{article}
\usepackage{asp2010}

\resetcounters

\begin{document}

\title{A possible solution to the mass-loss problem in M-type AGB stars}
\author{Hans Olofsson$^{1,2}$, Sofia Ramstedt$^3$, Stephane Sacuto$^4$, Matthias Maercker$^{3,5}$, Susanne H\"ofner$^6$, and Franz Kerschbaum$^4$
\affil{$^1$Onsala Space Observatory, Dept. of Radio and Space Science, Chalmers University of
Technology, SE-43992 Onsala, Sweden}
\affil{$^2$Department of Astronomy, SE-10691 Stockholm, Sweden}}
\affil{$^3$Argelander Institut f\"ur Astronomie, Auf dem H\"ugel 71, DE-53121 Bonn, Germany}
\affil{$^4$University of Vienna, Department of Astronomy, T{\"u}rkenschanzstra\ss{}e
17, A-1180 Vienna, Austria}
\affil{$^5$European Southern Observatory, Karl Schwarzschild Str. 2, Garching bei M\"unchen, Germany}
\affil{$^6$Department of Physics and Astronomy, Uppsala University, Sweden}

\begin{abstract}
Mass loss is a fundamental, observationally well-established feature of
AGB stars but many aspects of this process still remain to be understood. To date, self-consistent dynamical models of dust-driven winds reproducing the observed mass-loss rates seem only possible for M-type stars if the grains in the close circumstellar environment grow to larger sizes than previously assumed. In order to study the grain-size distribution where the mass loss is initiated, high-spatial-resolution interferometry observations are necessary. We have observed two M-type stars using the VLTI/MIDI instrument to constrain the dust-grain sizes through modeling the 10\,$\mu$m silicate feature. Complementary observations are scheduled and we will present preliminary results.
\end{abstract}

\section{Background}
The inclusion of frequency-dependent radiative transfer in hydrodynamical models revealed difficulties with driving the wind at the observed mass-loss rates in S- and M-type AGB stars \citep{woit06}. Recent work by \citet{hofn08} has shown that this can be solved when conditions in the upper atmosphere allow iron-free silicate grains to grow to larger sizes than in the previous models ($>$\,0.1$\mu$m). For grains large enough, scattering effects become important enough to dominate the driving and mass-loss rates of the observed magnitudes are reproduced.

\section{Observations}
To test this scenario the Very Large Telescope Interferometer (VLTI) of ESO's Paranal Observatory was used with the MID-infrared Interferometric recombiner (MIDI) \citep{leinetal03}. MIDI combines the light from two telescopes providing interferometric measurements in the N-band atmospheric window. We have observed two sources (RT Vir and R Crt) using the 130, 60, and 30 m baselines and obtained high-resolution (R=230) correlated spectra covering the 10\,$\mu$m silicate feature. With MIDI/VLTI we can probe how the spectral appearance of the emission changes from the
extended dust-free atmosphere to the inner dusty region. The shape of the 10\,$\mu$m silicate feature puts constraints on the size of the dust grains and by modeling the emission the fraction of large grains can be constrained together with the scattering scenario.

\begin{figure}[htbp]
\begin{center}
\includegraphics[width=\columnwidth]{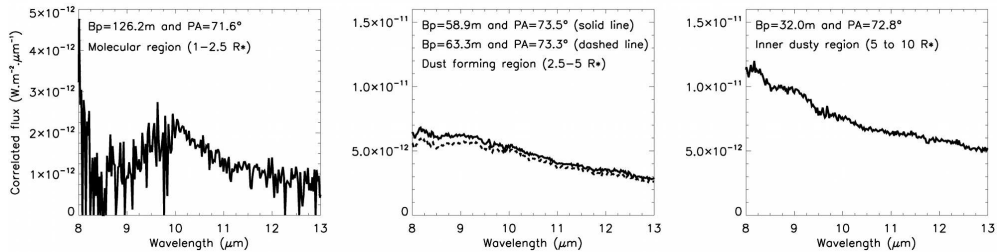}
\caption{The correlated flux of RT~Vir as measured at the three different baselines. The left plot shows the results using the longest ($\approx$130\,m) baseline. This probes 1-2.5 stellar radii and corresponds to the molecular region inside the dust formation zone. The middle plot shows the results using the intermediate ($\approx$60\,m) baselines. This probes 2.5-5 stellar radii and corresponds to the dust formation zone. The right plot shows the results using the shortest ($\approx$30\,m) baseline. This probes 5-10 stellar radii and corresponds to the inner dusty region.}
\label{fig}
\end{center}
\end{figure}

So-far we have observed one of the sources (RT~Vir) using three different baselines (Fig.~\ref{fig}), and the other (R~Crt) using two. Further observations using more baselines have been scheduled together with additional observations to obtain a better S/N in the longest baseline.

\section{Results and future prospects}
Already in the current data set, a clear change in the spectral appearance going from longer to shorter baselines is apparent. Using the data at hand we have already started tests to constrain the molecular layer probed by the longest baseline. The aim is to make sure that there is no confusion between molecular and dust emission. The dust emission will then be modeled in order to determine the dust composition and grain size distribution with the final goal of evaluating the large-grain scenario.

\acknowledgements SR acknowledges support by the Deutsche Forschungsgemeinschaft (DFG) through the Emmy Noether Research grant VL 61/3-1.

\bibliography{olofsson}

\end{document}